\documentclass[%
 reprint,
superscriptaddress,
nofootinbib,
 amsmath,
 amssymb,
 aps,
 prx,
 longbibliography,
]{revtex4-1}

\usepackage{graphicx}%
\usepackage{dcolumn}%
\usepackage{bm}%

\usepackage[a4paper,lmargin={2.2cm},rmargin={2.2cm},tmargin={2.2cm},bmargin={2.2cm}]{geometry}

\usepackage{natbib}
\usepackage{amsfonts}
\usepackage{amsmath}
\usepackage{amssymb}
\usepackage{graphicx} %
\usepackage{caption}
\usepackage{subcaption} %
\usepackage{booktabs} %
\usepackage{tikz}
\usepackage{lipsum}
\usepackage{xargs} %
\usepackage[
    separate-uncertainty=true, 
    round-mode=uncertainty,
    round-precision=1,
]{siunitx} %
\usepackage[export]{adjustbox}  %
\usepackage{placeins}
\usepackage{hyperref}
\usepackage{subfiles}
\usepackage[normalem]{ulem}
\usepackage{ragged2e}
\usepackage[colorinlistoftodos,prependcaption,textsize=tiny]{todonotes}

\definecolor{linkcolor}{HTML}{0b5394}

\hypersetup{
    colorlinks=true,
    linkcolor=linkcolor,     %
    citecolor=linkcolor,    %
    urlcolor=linkcolor       %
}

\usepackage{array,mathtools,amssymb,booktabs}
\newcolumntype{C}{>{$}c<{$}}

\AtBeginDocument{%
  \heavyrulewidth=.08em
  \lightrulewidth=.05em
  \cmidrulewidth=.03em
  \belowrulesep=.65ex
  \belowbottomsep=0pt
  \aboverulesep=.4ex
  \abovetopsep=0pt
  \cmidrulesep=\doublerulesep
  \cmidrulekern=.5em
  \defaultaddspace=.5em
}
\newcommand{\etarel}{$\eta^{\text{rel}}$}
\newcommand{\ptrel}{$p_\mathrm{T}^{\text{rel}}$}
\newcommand{\phirel}{$\phi^{\text{rel}}$}

\newcommand{\pt}{$p_\text{T}$}
\newcommand{\iptrans}{$d_0$}
\newcommand{\iptranserr}{$\sigma_{d_0}$}
\newcommand{\iplong}{$d_z$}
\newcommand{\iplongerr}{$\sigma_{d_z}$}
\newcommand{\jetclass}{\textsc{JetClass}}

\newcommand{\qcd}{$q/g$}
\newcommand{\zqq}{$Z\to q\bar{q}$}
\newcommand{\wqq}{$W\to qq'$}
\newcommand{\thad}{$t\to bqq'$}
\newcommand{\tlep}{$t\to b\ell\nu$}
\newcommand{\hbb}{$H\to b\bar{b}$}
\newcommand{\hcc}{$H\to c\bar{c}$}
\newcommand{\hgg}{$H\to gg$}
\newcommand{\hqqqq}{$H\to 4q$}
\newcommand{\hlnuqq}{$H\to \ell\nu qq'$}

\definecolor{goodmagenta}{HTML}{AA3377}
\definecolor{goodgreen}{HTML}{228833}
\definecolor{goodblue}{HTML}{4477AA}
\definecolor{goodyellow}{HTML}{CCBB44}
\definecolor{goodpink}{HTML}{EE6677}
\definecolor{goodgrey}{HTML}{BBBBBB}
\definecolor{pink}{HTML}{E16EAB}

\newcommandx{\commentjosch}[1][]{\textcolor{blue}{JB: #1}}
\newcommandx{\commentgregor}[1][]{\textcolor{violet}{GK: #1}}

\newcommandx{\todotikzpicture}[2][1=]{%
  \begin{tikzpicture}
    \node at (0,0) {#2};
    \node[text=red, align=left, fill=white, fill opacity=0.9, text opacity=1] at (0, 0) {TODO:\\#1};
  \end{tikzpicture}
}
\newcommandx{\todoNote}[2][1=]{\todo[linecolor=goodyellow,backgroundcolor=goodyellow!25,bordercolor=goodyellow,#1]{\sffamily#2}}
\newcommandx{\todoUnsure}[2][1=]{\todo[linecolor=goodmagenta,backgroundcolor=goodmagenta!15,bordercolor=goodmagenta,#1]{\sffamily\textit{Unsure:}\\#2}}
\newcommandx{\todoChange}[2][1=]{\todo[linecolor=orange,backgroundcolor=orange!25,bordercolor=orange,#1]{\sffamily\textit{Change:}\\#2}}
\newcommandx{\todoAdd}[2][1=]{\todo[linecolor=goodyellow,backgroundcolor=goodyellow!25,bordercolor=goodyellow,#1]{\sffamily\textit{Add:}\\#2}}
\newcommandx{\todoInfo}[2][1=]{\todo[linecolor=goodgreen,backgroundcolor=goodgreen!25,bordercolor=goodgreen,#1]{\sffamily\textit{Info:}\\#2}}
\newcommandx{\todoImprovement}[2][1=]{\todo[linecolor=goodblue,backgroundcolor=goodblue!25,bordercolor=goodblue,#1]{\sffamily\textit{Improvement:}\\#2}}
\newcommandx{\todoCheck}[2][1=]{\todo[linecolor=goodgrey,backgroundcolor=goodgrey!25,bordercolor=goodgrey,#1]{\sffamily\textit{Check:}\\#2}}

\begin{document}

\title{Flow Matching Beyond Kinematics: Generating Jets with Particle-ID and Trajectory Displacement Information}

\author{Joschka Birk}
\email{joschka.birk@uni-hamburg.de}
\affiliation{
    Institute for Experimental Physics, Universität Hamburg \\
    Luruper Chaussee 149, 22761 Hamburg, Germany
}

\author{Erik Buhmann}
\affiliation{
    Institute for Experimental Physics, Universität Hamburg \\
    Luruper Chaussee 149, 22761 Hamburg, Germany
}

\author{Cedric Ewen}
\affiliation{
    Institute for Experimental Physics, Universität Hamburg \\
    Luruper Chaussee 149, 22761 Hamburg, Germany
}

\author{Gregor Kasieczka}%
\affiliation{
    Institute for Experimental Physics, Universität Hamburg \\
    Luruper Chaussee 149, 22761 Hamburg, Germany
}

\author{David Shih}
\affiliation{
    New High Energy Theory Center, Rutgers University\\
    Piscataway, New Jersey 08854-8019, USA
}

\begin{abstract}

We introduce the first generative model trained on the \jetclass\ dataset. Our model generates jets at the constituent level, and it is a permutation-equivariant continuous normalizing flow (CNF) trained with the flow matching technique. It is conditioned on the jet type, so that a single model can be used to generate the ten different jet types of \jetclass. For the first time, we also introduce a generative model that goes beyond the kinematic features of jet constituents. The \jetclass\ dataset includes more features, such as particle-ID and track impact parameter, and we demonstrate that our CNF can accurately model all of these additional features as well. Our generative model for \jetclass\ expands on the versatility of existing jet generation techniques, enhancing their potential utility in high-energy physics research, and offering a more comprehensive understanding of the generated jets.

\end{abstract}

\maketitle

\section{Introduction}

Recently there has been considerable interest and activity in generative
modeling for jet constituents. While showering and hadronization with standard
programs such as \textsc{Pythia}~\cite{pythia8} and \textsc{Herwig}~\cite{herwig2008}
are not a major computational bottleneck at
the LHC~\cite{Evans:2008zzb}, learning the properties of jets from data still has interesting potential applications.
For example, generative modeling at the jet
constituent level can be used to improve the performance of anomaly 
detection~\cite{full_phase_space_anomaly} techniques, since this allows to cover
the full phase space instead of using solely high-level features that are calculated 
from the jet constituents. 

More generally,  learning jets is 
an interesting laboratory for method development. In particular, it has
been fruitful and effective to view the jet constituents as a high-dimensional
point cloud, and to devise methods for point cloud generative models that
incorporate permutation invariance. This route has led to a number of 
state-of-the-art approaches, recently explored in~\cite{mpgan2022, käch2022jetflow, EPiC-GAN, pcjedi2023, FPCD, käch2023attention_MDMA, CaloClouds, PCDroid, CaloClouds_2, epicly}, that combine
different permutation-invariant layers such as transformers~\cite{AttentionIsAllYouNeed} and the
EPiC~(Equivariant Point Cloud) layer~\cite{EPiC-GAN}, with state-of-the-art generative modeling frameworks such
as diffusion~\cite{sohldickstein2015deep, song2020generative_estimatingGradients, song2020improved_technieques_for_sorebased_geneerative, ho2020denoising, song2021scorebased_generativemodelling} 
and flow matching~\cite{liu2022flow_flowmatching1,albergo2023building_flowmatching2,lipman2023flow,tong2023improvingflowmatching}. 
Successful models developed for jet point
clouds can also potentially be adapted to other important point cloud
generative modeling problems such as for fast emulation of GEANT4~\cite{Geant4_1,Geant4_2,Geant4_3} calorimeter
showers~\cite{CaloClouds, CaloClouds_2, acosta2023comparison,scham2023deeptreegan}.
Finally, while event generation with generative models
has concentrated primarily on low multiplicities and fixed 
structures~\cite{EventGen6,EventGen2,EventGen3,EventGen1,EventGen5,EventGen7}, 
recent, permutation invariant approaches that allow for dynamic multiplicities exist
as well~\cite{JetGPT,Butter:2023ira}.

\begin{figure}[t]
    \centering
    \includegraphics[width=\linewidth]{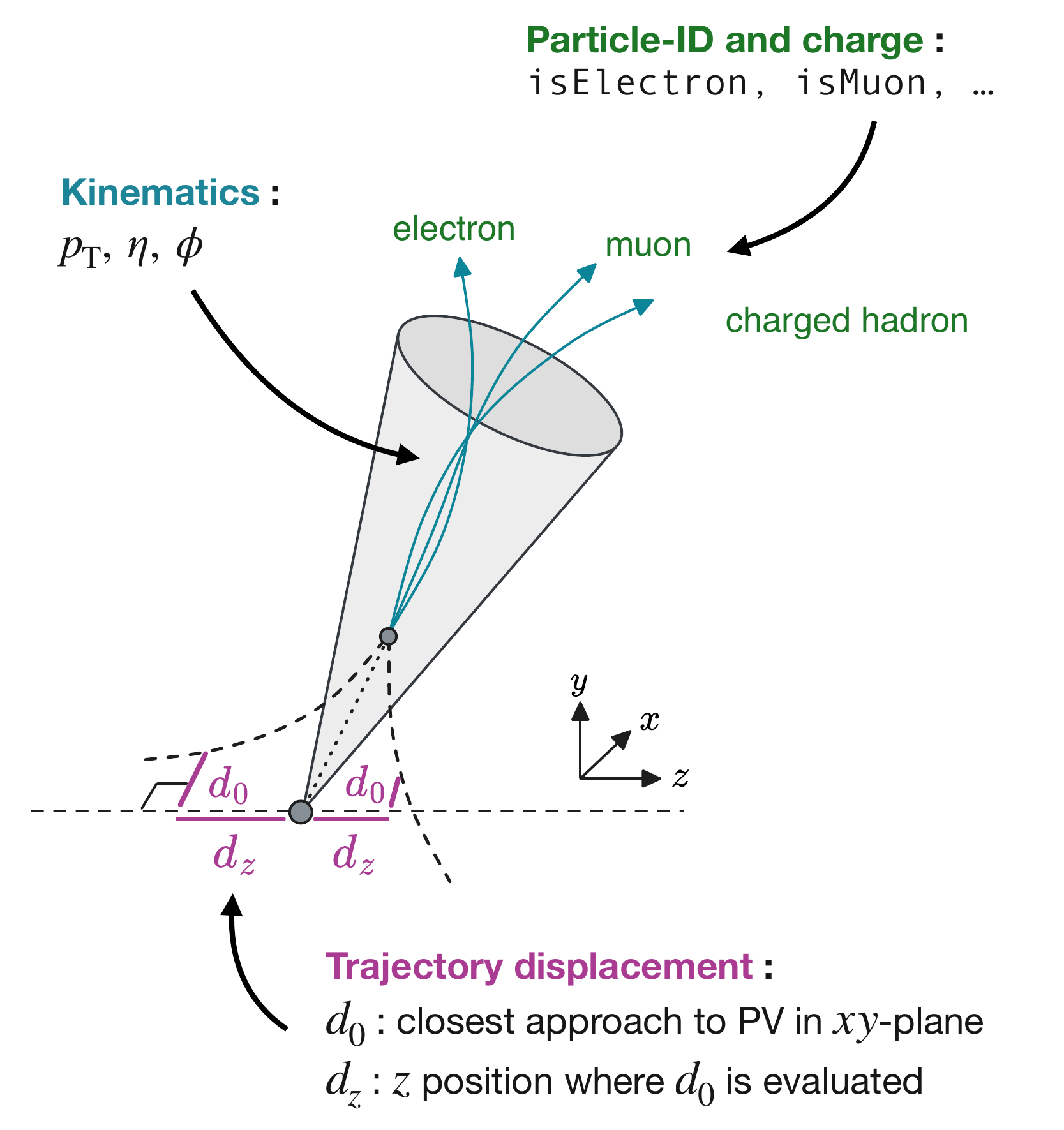}
    \caption{
        Schematic overview of the different jet constituent features
        available in the \jetclass\ dataset.
        The horizontal line at the bottom represents the beam axis
        and the circle on this line represents the primary vertex (PV).
    }
    \label{fig:beyond_kinematics_sketch}
\end{figure}

\begin{figure*}
    \centering
    \begin{subfigure}[t]{0.49\textwidth}
        \includegraphics[height=0.58\textwidth]{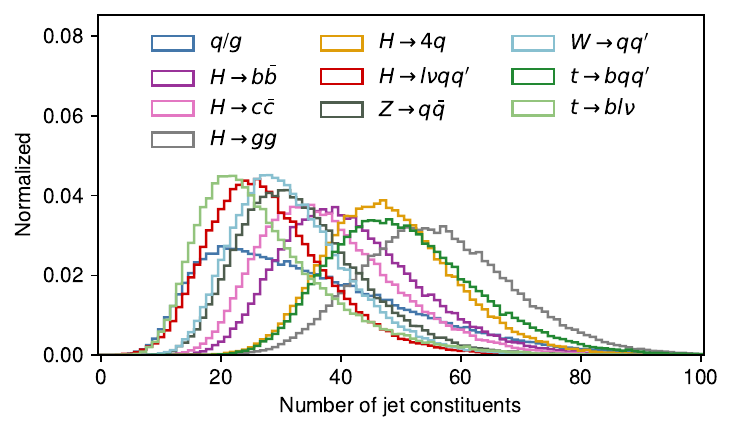}
        \subcaption{}
        \label{fig:jetclass_features_overview_jet_nparticles}
    \end{subfigure}
    \begin{subfigure}[t]{0.49\textwidth}
        \includegraphics[height=0.59\textwidth]{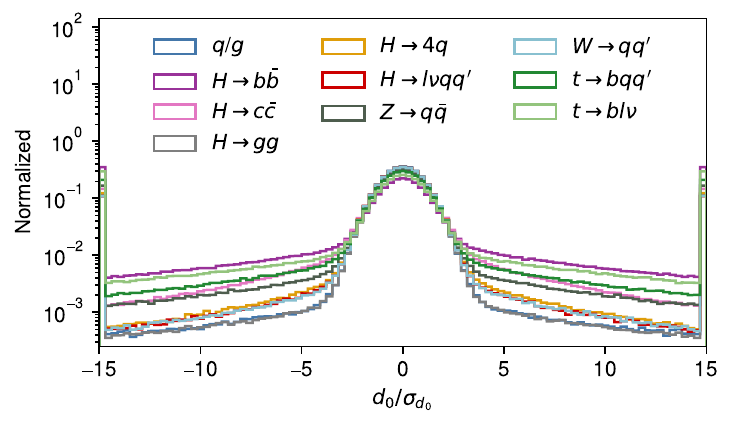}
        \subcaption{}
        \label{fig:jetclass_features_overview_part_sd0}
    \end{subfigure}
    \begin{subfigure}[t]{0.49\textwidth}
        \includegraphics[height=0.57\textwidth]{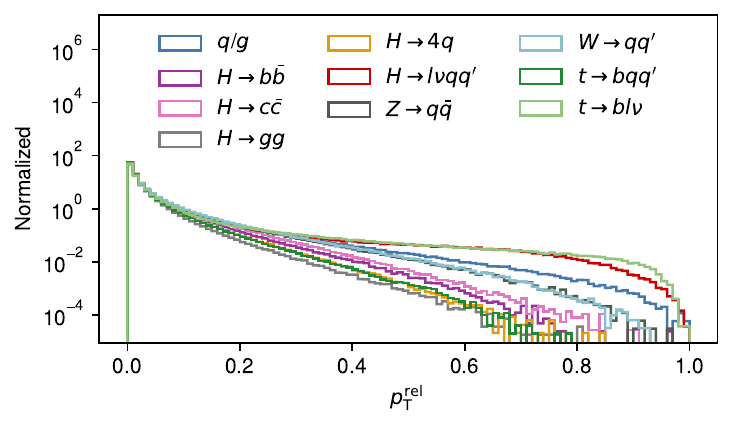}
        \subcaption{}
        \label{fig:jetclass_features_overview_part_ptrel}
    \end{subfigure}
    \hfill
    \begin{subfigure}[t]{0.49\textwidth}
        \includegraphics[height=0.58\textwidth]{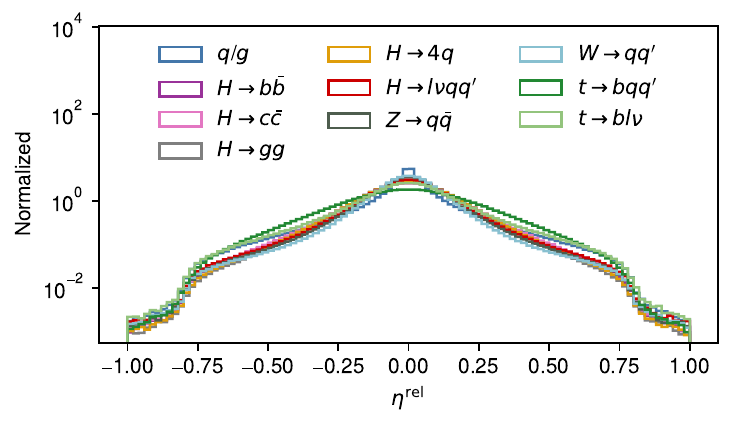}
        \subcaption{}
        \label{fig:jetclass_features_overview_part_etarel}
    \end{subfigure}
    \caption{
        Overview of some of the features from the \jetclass\ dataset:
        \textbf{(a)} shows the number of jet constituents, 
        \textbf{(b)} shows the significance of the transverse impact
        parameter $d_0$,
        \textbf{(c)} shows the fraction of the jet \pt\ carried by
        the jet constituent and 
        \textbf{(d)} shows the difference \etarel\ between the constituent
        pseudorapidity and the jet axis.
        The impact parameter significance $d_0 / \sigma_{d_0}$ is only shown for 
        charged particles since the impact parameter is 0 for neutral particles.
        The number of jet constituents is a jet-level feature our model is 
        conditioned on while the remaining features correspond to constituent-level
        features that our model generates.
    }
    \label{fig:jetclass_features_overview}
\end{figure*}

So far, efforts for jet generation have focused almost exclusively on the \textsc{JetNet} dataset of
Refs.~\cite{JetNet,JetNet150}. Originally generated by~\cite{Coleman:2017fiq}, this
dataset was subsequently adopted in the works of Ref.~\cite{mpgan2022} as a 
benchmark dataset for jet generative modeling. However, the \textsc{JetNet} dataset has
a number of drawbacks that are readily becoming apparent. First, its  limited size (180k jets per jet type) means
there are not enough jets in \textsc{JetNet}
to facilitate the training of state-of-the-art generative models as well as metrics
such as the binary classifier metric~\cite{CaloFlowKrause_2023,understandLimitationsOfGen} which require additional training data.
Second, \textsc{JetNet} uses small-radius ($R=0.4$) jets (although the description in \cite{mpgan2022} incorrectly states a cone-size of $R=0.8$ which is in disagreement with the observed angular distribution of constituents). 
This can lead to decay products not being
fully contained in the jet, which can be seen e.g. in distributions such 
as the jet mass for top quarks, where there is a prominent secondary mass peak
(as can be seen e.g. in~\cite{epicly},~Fig.\,4).
Finally, \textsc{JetNet} focuses solely on the kinematics of the jet
constituents, whereas there is a wealth of additional information inside the
jets that could also be modeled, such as trajectory displacement, charge, and
particle ID as illustrated in \autoref{fig:beyond_kinematics_sketch}. 

In this work, we introduce the first generative model for jet constituents trained on the larger and more substantial
\jetclass\ dataset~\cite{JetClass}. 
Other than demonstrating that existing techniques scale
well to this new dataset, we also tackle new challenges introduced by the
\jetclass\ dataset, including the additional features mentioned above. We show
that a single architecture consisting of a continuous normalizing flow (CNF)~\cite{chen2019neural_CNF} constructed out of EPiC layers and trained with the flow matching technique~\cite{liu2022flow_flowmatching1,albergo2023building_flowmatching2,lipman2023flow,tong2023improvingflowmatching} can learn to model all ten jet
types and all features of \jetclass\ well, both kinematics and
beyond. 
Using the much larger statistics provided by \jetclass\, we also train binary
classifier metrics on the generative model for the first time, which 
allows us to further evaluate our results.

The rest of this paper is structured as follows: ~\autoref{sec:dataset} describes 
the \jetclass\ dataset in more detail; 
~\autoref{sec:model} introduces the used generative model; 
\autoref{sec:results} provides an overview of the generative performance; 
and finally \autoref{sec:conclusion} concludes this work.

\section{Dataset}

\label{sec:dataset}

We use the \jetclass\ dataset~\cite{JetClass}, originally introduced in Ref.~\cite{ParT}. 
This dataset encompasses various features both on jet level and on jet constituent level
for ten different types of jets initiated by gluons and quarks ($q/g$),
top quarks ($t$), as well as $W$, $Z$, and $H$ bosons.
Jets initiated by a top quark or a Higgs boson are further categorized based on
their different decay channels, resulting in the following ten categories:
\qcd, \thad, \tlep, \zqq, \wqq, \hbb, \hcc, \hgg, \hqqqq, and \hlnuqq.

\begin{table*}
    \centering
    \caption{
        Features used for the studies in this paper. 
        All features are either taken directly from the \jetclass\ dataset or
        calculated from existing entries in the \jetclass\ dataset.
        The model is conditioned
        on the jet features (first block) during generation. The remaining features are the constituent
        features our model is trained to generate.
    }
    \label{tab:features_definition}
    \begin{tabular}{c c p{0.65\textwidth}}
        Category & Variable & Definition \\
        \midrule
        \midrule
        & Jet type & Indicating the type of the jet (i.e. \thad, \qcd, ...) \\
        Jet features & $n_\mathrm{const}$ & Number of constituents in the jet \\
         & $p_\mathrm{T}^\mathrm{jet}$ & Transverse momentum of the jet \\
        & $\eta^\mathrm{jet}$ & Pseudorapidity of the jet \\
        \midrule
        \midrule
        & \etarel & Difference in pseudorapidity $\eta$ between the constituent and the jet axis \\
        Kinematics & \phirel & Difference in azimuthal angle $\phi$ between the constituent and the jet axis \\
        & \ptrel & Fraction of the constituent \pt\ and the jet \pt \\
        \midrule
        & \iptrans & Transverse impact parameter value \\
        Trajectory & \iplong & Longitudinal impact parameter value \\
        displacement & \iptranserr & Error of measured transverse impact parameter \\
        & \iplongerr & Error of measured longitudinal impact parameter \\
        \midrule
        & charge & Electric charge of the particle \\
        & isChargedHadron & Flag if the particle is a charged hadron (\texttt{|pid|==211 or 321 or 2212}) \\
        Particle & isNeutralHadron & Flag if the particle is a neutral hadron (\texttt{|pid|==130 or 2112 or 0}) \\
        identification & isPhoton & Flag if the particle is a photon (\texttt{pid==22}) \\
        & isElectron & Flag if the particle is an electron (\texttt{|pid|==11}) \\
        & isMuon & Flag  if the particle is a muon (\texttt{|pid|==13}) \\
        \midrule
  \end{tabular}
\end{table*}

The jets are extracted from simulated events that are generated with
\textsc{MadGraph5\_aMC@NLO}~\cite{madgraph}. Parton showering and hadronization
is performed with \textsc{Pythia}~\cite{pythia8} and detector effects are
simulated with \textsc{Delphes}~\cite{delphes} using the CMS detector~\cite{CMS_experiment} card.
The jets are clustered with the anti-$k_\mathrm{T}$ algorithm~\cite{antikt} with
a jet radius of $R=0.8$. %
Jets are required to have a transverse momentum of
\mbox{$ \SI{500}{GeV} < p_\mathrm{T}^\mathrm{jet} < \SI{1000}{GeV}$} and a pseudorapidity of
\mbox{$|\eta^\mathrm{jet}| < 2$}.
Furthermore, for all jets, except for the \qcd\ jets, only those that 
contain all decay products of the boson or top
quark are considered in the dataset.

Representative distributions of some of the features available in the \jetclass\ dataset are
presented in \autoref{fig:jetclass_features_overview}, illustrating substantial
distinctions among the various jet categories.
As can be seen in \autoref{fig:jetclass_features_overview_jet_nparticles},
the dataset covers jet types with a large variety of number of jet constituents.
Additionally, \autoref{fig:jetclass_features_overview_part_sd0} illustrates the expected 
tails of varying extents in the trajectory displacement.
Notably, jet categories that do not include $b$ quarks or $c$ quarks have a rather narrow
peak around 0, while jet categories that do include those types of quarks have a
more spread out distribution, corresponding to the long lifetime of $b$- and $c$-hadrons.

The features that are used by our model are listed with 
the corresponding definitions in \autoref{tab:features_definition}.
By conditioning our model on the jet type, we can train one model that is
capable of generating all ten jet types.
Furthermore, the jet transverse momentum $p_\mathrm{T}^\mathrm{jet}$ and the jet 
pseudorapidity $\eta^\mathrm{jet}$ are used as conditioning features
in order to be able to obtain the non-relative constituent-level features after
jet generation.

The kinematic features of the jet constituents are measured relative to
the jet axis. i.e.
\begin{align}
    \eta^\mathrm{rel} &= \eta^\mathrm{const} - \eta^\mathrm{jet} \\
    \phi^\mathrm{rel} &= \phi^\mathrm{const} - \phi^\mathrm{jet} \\
    p_\mathrm{T}^\mathrm{rel} &= p_\mathrm{T}^\mathrm{const} / p_\mathrm{T}^\mathrm{jet} \,.
\end{align}
The trajectory displacement features as well as the discrete particle identification
features are taken directly from the \jetclass\ dataset.
The jet constituents are assumed to be massless, as we do not include the
particle energy in our model.
In total, our model thus utilizes 13 constituent features to comprehensively
characterize the constituents within generated jets.
This set of features represents all available constituent features, 
from the \jetclass\ dataset, with the only exception being the 
constituent energy.

\section{Model}

\label{sec:model}

Overall, we follow the recent flow matching paradigm~\cite{liu2022flow_flowmatching1,albergo2023building_flowmatching2,lipman2023flow,tong2023improvingflowmatching}, using Gaussian probability paths as explained in Ref.~\cite{epicly}.

The architecture of our model is based on the EPiC-FM~(Equivariant Point Cloud Flow Matching) model which was 
introduced in Ref.~\cite{epicly}.
It uses EPiC 
layers \cite{EPiC-GAN} to transform jet constituents as point clouds while preserving permutation symmetry.
Besides the point cloud, EPiC layers also take as input a global context vector, 
which learns to encode higher-level information 
about the whole point cloud by using both mean and sum aggregation over the points.
The use of the global context vector is responsible for the computational efficiency of the EPiC layers -- unlike the quadratic scaling for architectures that include edge features between all pairs of points (such as transformers), the computational cost of EPiC layers only scales linearly with
the number of points.
\begin{figure*}
    \includegraphics[width=0.99\linewidth]{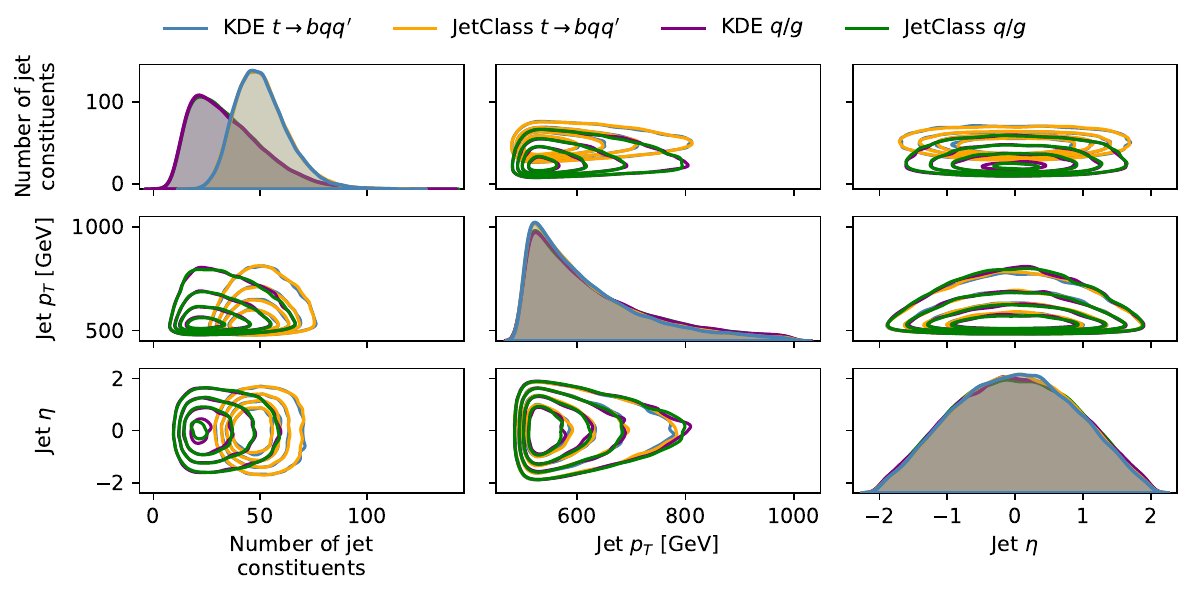}
    \caption{
        Pair plot of the jet features of the \jetclass\ dataset and the KDE samples
        for the \thad{} and \qcd{} jet types.
        The features generated by the KDE are used as conditioning features for the
        final generative model.
        The diagonal shows the distribution of the individual features, while the
        off-diagonal plots show the correlation between the features.
        The lines in the off-diagonal plots correspond to iso-proportions 
        (20\%, 40\%, 60\% and 80\%) of the density (e.g. 20\% of the data points
        are outside the 20\% line). Each of the 4 plotted datasets contains 100k jets.
    }
    \label{fig:kde_verification_pairplot}
\end{figure*}
Compared to the model in Ref.~\cite{epicly}, the hyperparameters were adapted to obtain 
good performance with the increased complexity of the \jetclass\ dataset:
the number of EPiC layers was increased from 6 to 20, the
embedding dimension was increased from 128 to 300, and the global vector
dimension was increased from 10 to 16.
A full list of the specific hyperparameters of the model is 
summarized in Appendix \ref{sec:appendix_hyperparameters}, \autoref{tab:baseline_model_hyperparameters}.

The jet-type conditioning is realized with a one-hot encoding, while the 
jet transverse momentum $p_\mathrm{T}^\mathrm{jet}$ and jet 
pseudorapidity $\eta^\mathrm{jet}$ are used without any preprocessing.
For jet generation, we fit a kernel density estimator (KDE) to the jet features
($p_\mathrm{T}^\mathrm{jet}$, $\eta^\mathrm{jet}$, and $n_\mathrm{const}$)
for each jet type individually, and sample 
from the KDE to obtain the conditioning features and the number of constituents
for the generated jets.
Since the KDE samples are continuous, we round the
number of constituents to the nearest integer and clip all three features to
the minimum and maximum values found in the training dataset.
A verification that the KDE samples are representative
of the \jetclass\ dataset is shown in \autoref{fig:kde_verification_pairplot}.
For each jet, our model acts on a tensor of shape $(N, C)$, with 
$N=128$ being the maximum number of jet constituents and $C$ being the 
number of constituent features. To generate jets with fewer constituents ($n_\mathrm{const}<128$), we apply a binary mask throughout training; this ensures that the inputs and outputs of the EPiC layers are valid and in particular guarantees that the mean and sum aggregation steps within the EPiC layers are correctly normalized.
The particle features are shifted to mean 0 and scaled to the same standard deviation.
Furthermore, only particles with $|\eta^\text{rel}| < 1$ are considered to remove a few
particles in the tails of this distribution.\footnote{
    The jets in the \jetclass\ dataset are reported to correspond to 
    a jet radius of $R=0.8$, but some jet constituents have 
    $|\eta^\text{rel}| > 0.8$. 
    However, most of the constituents fall into $|\eta^\text{rel}| \leq 0.8$, 
    so only a very small fraction of jet constituents is removed by this cut.
}

The model is trained for 500 epochs with a training sample size of 3M jets (\num{300 000} jets
per jet type).\footnote{
    It should be noted that this is just a small fraction 
    (\SI{3}{\percent}) of the training jets available in the \jetclass\ dataset. 
    Training with 5M instead of 3M jets did not show significant improvements in
    some preliminary trials. Given this, and the large potential computing
    expense from scaling up, we limited the training set size to this small
    fraction of \jetclass\ in this first study.
    We still benefited from the larger size of 
    \jetclass\ in that it enabled us to train the binary classifier metrics.
}

The AdamW optimizer~\cite{AdamW} is used with a maximum learning rate of 0.001
and weight decay of \num{5e-5}.
The learning rate is scheduled with a cosine-annealing learning rate scheduler 
with a linear warm-up period over 20 epochs.
The model after the last epoch is chosen for evaluation.\footnote{
    Given that the training of this model is very stable, the last 
    epoch essentially also corresponds to the epoch with the lowest loss.
    Thus, for simplicity the model from the last epoch was chosen.
}
For the jet generation, we use the midpoint ODE solver, a 2\textsuperscript{nd} 
order Runge-Kutta variant.
After jet generation, the \ptrel\ values of the generated jet constituents 
are clipped to the minimum
and maximum values that are found in the training dataset.
Furthermore, the electric charge is rounded to the nearest
integer within \mbox{\{-1, 0, 1\}}. The remaining discrete particle-identification features are
post-processed with an argmax operation such that the 
particle-ID feature with the maximum value is set to
1 and the remaining ones are set to 0. This ensures that generated
particles are unambiguously assigned to one particle type.
It should be noted that this might not be the ideal handling of such discrete 
features, which could be improved and further investigated in future work.
The model is implemented in \textsc{PyTorch}~\cite{pytorch_NEURIPS2019_9015}
using \textsc{PyTorch Lightning}~\cite{Falcon_PyTorch_Lightning_2019}.

\section{Results}

\label{sec:results}

In the following, we first evaluate the distributions of the 
constituent-level features that the model was trained on.
Afterwards, we evaluate jet-level features that are derived from the constituent-level
features, namely the jet mass, the $N$-subjettiness~\cite{Thaler_2011_subjettiness} ratios\footnote{
   For the calculation of the subjettiness, the particles are clustered into two and three subjets
   with the \mbox{$k_t$-algorithm} and the winner-takes-all clustering scheme using 
   \texttt{scikit-hep/fastjet}~\cite{aryan_roy_2023_7504167}.
}
$\tau_{32}$ and $\tau_{21}$ as well as the $D_2$~\cite{larkoski2016analytic} observable,
which is a ratio of energy correlation functions~\cite{Larkoski_2013_energy_correlation_functions}.
Lastly, we train a binary classifier as in Ref.~\cite{CaloFlowKrause_2023} to distinguish between real jets
from the \jetclass\ dataset and jets generated by our model,
which allows us to evaluate the quality of the generated jets further.

For the evaluation of our model, we generate 5M jets (i.e. \num{500 000} jets per jet type)
and evaluate the different feature categories separately.
Instead of using the Wasserstein-1 distance as a metric for
the agreement between the generated and the target distribution as in \cite{mpgan2022,EPiC-GAN,pcjedi2023}, we are following \cite{epicly} and using
the Kullback-Leibler (KL) divergence $\mathrm{KL}(p_{\text{JetClass}} || p_{\text{EPiC-FM}})$.
As explained there, this metric choice is motivated by the 
fact that the Wasserstein-1 distance is well-suited for detecting distributions with different support, but can lead to misleading results for cases 
where the distributions have the same overlapping support but differ mainly in their shapes.

\begin{figure*}
    \centering
    \begin{subfigure}[b]{0.49\textwidth}
        \includegraphics[height=5cm]{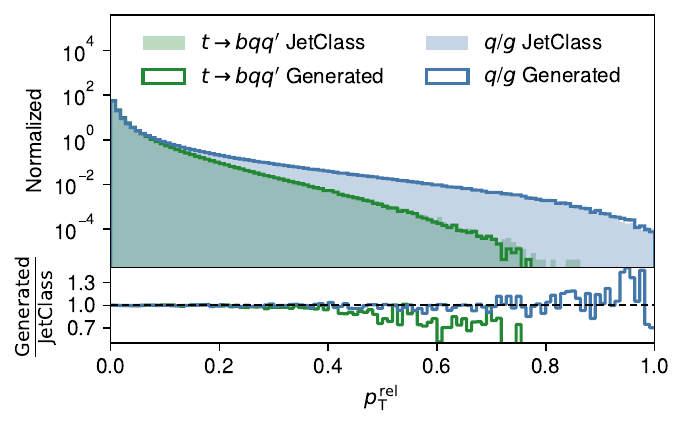}
        \subcaption{}
    \end{subfigure}
    \begin{subfigure}[b]{0.49\textwidth}
        \includegraphics[height=5cm]{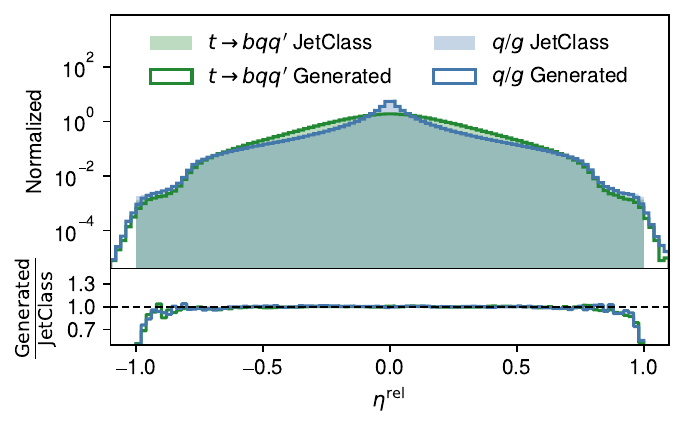}
        \subcaption{}
    \end{subfigure}
    \begin{subfigure}[t]{0.497\textwidth}
        \includegraphics[height=5cm]{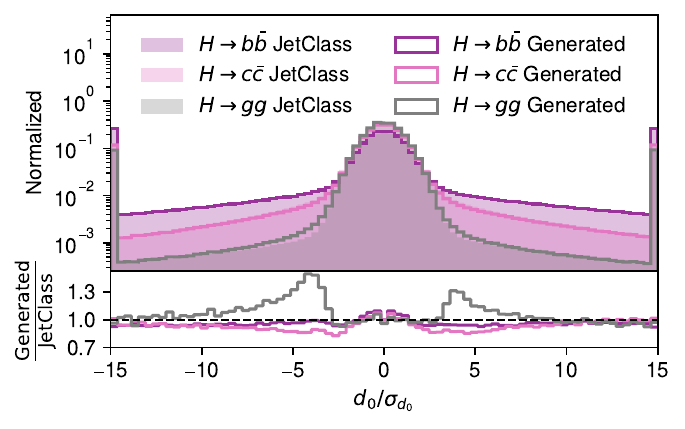}
        \subcaption{}
    \end{subfigure} \hfill
    \begin{subfigure}[t]{0.497\textwidth}
        \includegraphics[height=5cm]{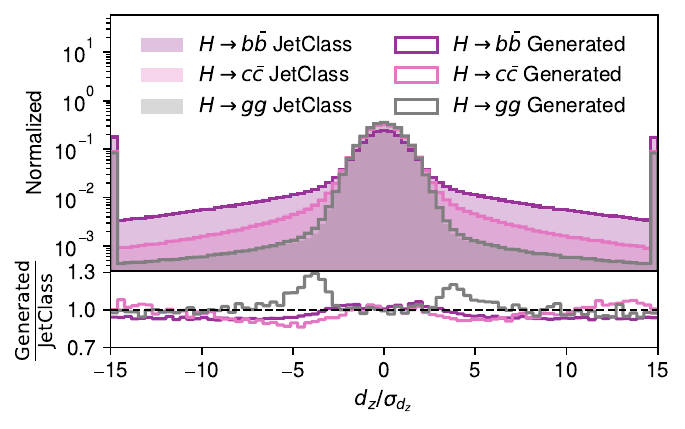}
        \subcaption{}
    \end{subfigure}
    \caption{
        Two kinematic features and the trajectory displacement
        of the jet constituents as generated by our model (solid lines) 
        in comparison to the distributions obtained from the \jetclass\ 
        dataset (semi-transparent histograms).
        The upper row shows \textbf{(a)} the relative transverse momentum
        and \textbf{(b)} the relative pseudorapidity.
        The lower row shows \textbf{(c)} the transverse impact parameter significance
        and \textbf{(d)} the longitudinal impact  parameter significance.
        Only charged particles are shown in the histograms of the
        trajectory displacement and the outermost bins show the underflow and overflow bins.
    }
    \label{fig:baseline_ip_and_kin_features}
\end{figure*}

The KL divergence values are calculated by splitting the generated
jets and the \jetclass\ jets into 10 batches of \num{50 000} jets each
and then calculate the KL divergence for each batch.
This allows us to evaluate the uncertainty of the KL divergence 
to be able to quantitatively compare it to the truth distribution.
The reported values are the mean of those 10 individual KL divergence values
and their standard deviation.
For the truth values of the KL divergence, we take two independent samples of the
test dataset and then calculate the KL divergence between those two samples
using the same procedure as when comparing the real and generated
jets.
The bin edges of the binned probability distributions are constructed
by calculating 100 equiprobable quantiles from the target distribution
and setting the edges of the outermost bins to the minimum and maximum
values found in the target distribution. 
Outliers of the distribution from the generated jets are put into the first and last
bin.

\subsection{Jet constituent modeling}

\begin{figure*}
    \centering
    \includegraphics[width=0.9\textwidth]{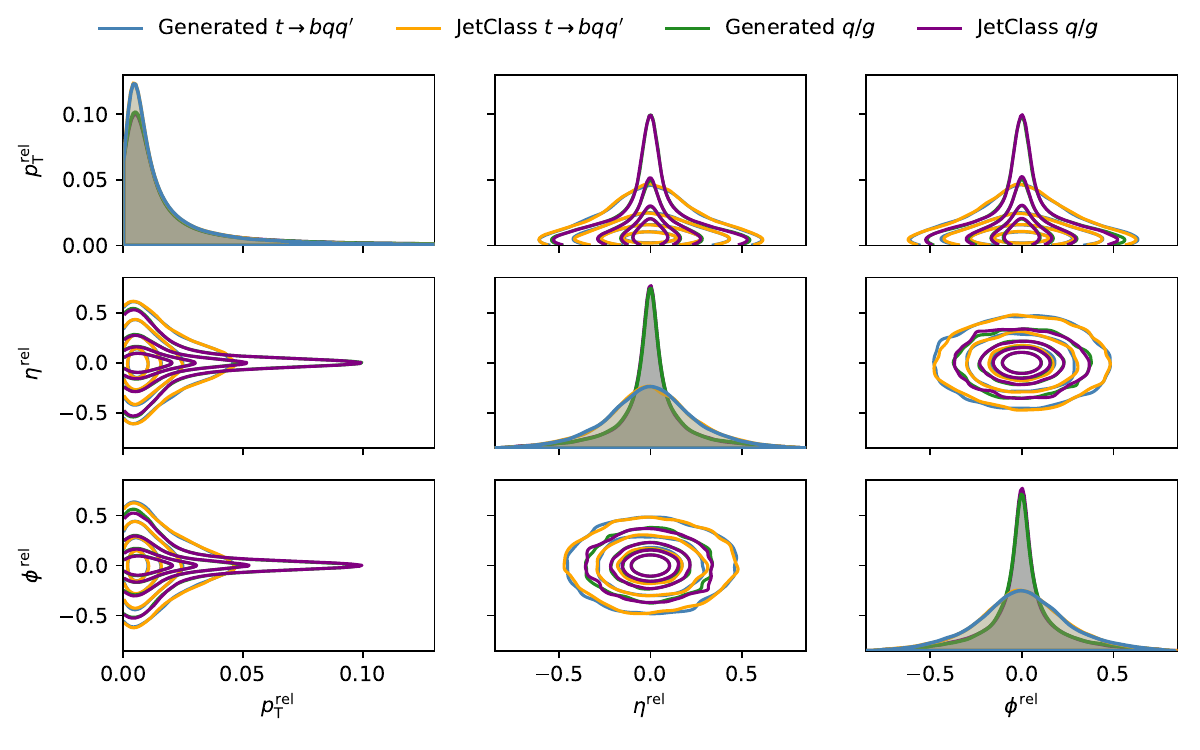}
    \caption{
        Pair plot of the kinematic features of the jet constituents for the
        \thad\ and \qcd\ jet types illustrating that the model correctly
        captures the correlations between the different kinematic features. 
        The lines in the off-diagonal plots correspond to iso-proportions 
        (20\%, 40\%, 60\% and 80\%) of the density (e.g. 20\% of the data points
        are outside the 20\% line). Each of the 4 plotted datasets contains 100k
        constituents.
    }
    \label{fig:baseline_kinematic_correlations}
\end{figure*}

A comparison of the kinematic features of the jet constituents is shown
in \autoref{fig:baseline_ip_and_kin_features}. For this comparison,
we chose to compare \thad\ jets and \qcd\ jets, since those two types
of jets have distinct characteristics in both the \etarel\ and the
\ptrel\ distribution. While particles in \thad\ jets have on average
a larger \etarel\ and thus have an overall wider \etarel\ distribution,
\qcd\ jets are more collimated, resulting in a sharper \etarel\ peak around
$\eta^\mathrm{rel}=0$.
Concerning the \ptrel\ distribution, \thad\ jets are
expected to show a smaller tail towards larger \ptrel\ values, since \thad\ jets
contain on average more constituents and thus the jet \pt\ is distributed
over more particles, leading to smaller \ptrel\ values.
The generated distributions of all three kinematic features agree very 
well with the corresponding distribution from the \jetclass\ dataset, showing
that our model is capable of generating jets of very different kinematic
properties. This is also confirmed by the KL divergence values listed
in \autoref{tab:baseline_model_kld_particle_level}, which only have a small
deviation from the truth values.
In addition to that, \autoref{fig:baseline_kinematic_correlations} shows a pair
plot of the kinematic features of the jet constituents, confirming that the
model accurately captures the correlations between those features.

\begin{table*}
    \centering
    \caption{
        Kullback-Leibler divergence of constituent-level features with respect 
        to the \jetclass\ dataset.
    }
    \label{tab:baseline_model_kld_particle_level}
\begin{tabular}{l|c|c|c|c|c}
&$\mathrm{KL}^{p_\mathrm{T}^\mathrm{rel}} ~ (\times 10^{-3})$ & $\mathrm{KL}^{\eta^\mathrm{rel}} ~ (\times 10^{-3})$ & $\mathrm{KL}^{\phi^\mathrm{rel}} ~ (\times 10^{-3})$ & $\mathrm{KL}^{d_0 / \sigma_{d_0}} ~ (\times 10^{-3})$ & $\mathrm{KL}^{d_z / \sigma_{d_z}} ~ (\times 10^{-3})$\\
\midrule
Truth (\qcd) & \num{0.05962 \pm 0.01034} & \num{0.08033 \pm 0.00970} & \num{0.07218 \pm 0.01001} & \num{0.13298 \pm 0.01427} & \num{0.11914 \pm 0.01577} \\
EPiC-FM (\qcd) & \num{0.07320 \pm 0.00713} & \num{0.13035 \pm 0.02168} & \num{0.11907 \pm 0.01153} & \num{2.22425 \pm 0.06336} & \num{3.90859 \pm 0.08540} \\
\midrule
Truth (\hbb) & \num{0.04807 \pm 0.00391} & \num{0.07479 \pm 0.01215} & \num{0.06021 \pm 0.00821} & \num{0.10227 \pm 0.01019} & \num{0.09546 \pm 0.01107} \\
EPiC-FM (\hbb) & \num{0.07047 \pm 0.00836} & \num{0.12195 \pm 0.02243} & \num{0.11699 \pm 0.02247} & \num{2.64120 \pm 0.08700} & \num{1.48675 \pm 0.07362} \\
\midrule
Truth (\hcc) & \num{0.05415 \pm 0.00703} & \num{0.07489 \pm 0.01378} & \num{0.08457 \pm 0.01507} & \num{0.10597 \pm 0.01915} & \num{0.10944 \pm 0.02522} \\
EPiC-FM (\hcc) & \num{0.07762 \pm 0.00683} & \num{0.11306 \pm 0.01941} & \num{0.13436 \pm 0.00885} & \num{2.81678 \pm 0.10350} & \num{2.03249 \pm 0.08109} \\
\midrule
Truth (\hgg) & \num{0.03438 \pm 0.00247} & \num{0.05780 \pm 0.01550} & \num{0.05272 \pm 0.00820} & \num{0.07463 \pm 0.01375} & \num{0.07582 \pm 0.00945} \\
EPiC-FM (\hgg) & \num{0.05684 \pm 0.00901} & \num{0.10059 \pm 0.01345} & \num{0.10103 \pm 0.01414} & \num{2.38777 \pm 0.09853} & \num{3.75019 \pm 0.13275} \\
\midrule
Truth (\hqqqq) & \num{0.03950 \pm 0.00466} & \num{0.05703 \pm 0.00719} & \num{0.05324 \pm 0.01082} & \num{0.08051 \pm 0.01142} & \num{0.08549 \pm 0.01939} \\
EPiC-FM (\hqqqq) & \num{0.05688 \pm 0.00626} & \num{0.11775 \pm 0.01993} & \num{0.10621 \pm 0.01515} & \num{2.30474 \pm 0.07151} & \num{3.37262 \pm 0.12772} \\
\midrule
Truth (\hlnuqq) & \num{0.07116 \pm 0.01293} & \num{0.09878 \pm 0.02302} & \num{0.09697 \pm 0.01441} & \num{0.13964 \pm 0.01103} & \num{0.13760 \pm 0.01663} \\
EPiC-FM (\hlnuqq) & \num{0.10317 \pm 0.01354} & \num{0.15427 \pm 0.02272} & \num{0.13721 \pm 0.00855} & \num{2.11118 \pm 0.07543} & \num{2.70495 \pm 0.09921} \\
\midrule
Truth (\zqq) & \num{0.06101 \pm 0.00819} & \num{0.07686 \pm 0.01193} & \num{0.07736 \pm 0.01304} & \num{0.13846 \pm 0.03211} & \num{0.13434 \pm 0.01236} \\
EPiC-FM (\zqq) & \num{0.07360 \pm 0.00982} & \num{0.14488 \pm 0.02458} & \num{0.12659 \pm 0.01858} & \num{3.26740 \pm 0.17081} & \num{2.17424 \pm 0.13619} \\
\midrule
Truth (\wqq) & \num{0.06281 \pm 0.01084} & \num{0.08797 \pm 0.00950} & \num{0.08648 \pm 0.00847} & \num{0.12368 \pm 0.01451} & \num{0.13684 \pm 0.01732} \\
EPiC-FM (\wqq) & \num{0.08246 \pm 0.00768} & \num{0.14039 \pm 0.02249} & \num{0.13022 \pm 0.01662} & \num{2.17685 \pm 0.04841} & \num{2.70335 \pm 0.11423} \\
\midrule
Truth (\thad) & \num{0.04260 \pm 0.00667} & \num{0.05973 \pm 0.00491} & \num{0.05715 \pm 0.00749} & \num{0.07989 \pm 0.01026} & \num{0.08392 \pm 0.00982} \\
EPiC-FM (\thad) & \num{0.08227 \pm 0.01122} & \num{0.08847 \pm 0.01358} & \num{0.09984 \pm 0.01072} & \num{1.84127 \pm 0.06797} & \num{1.86226 \pm 0.10431} \\
\midrule
Truth (\tlep) & \num{0.07229 \pm 0.01400} & \num{0.12652 \pm 0.01816} & \num{0.14858 \pm 0.03336} & \num{0.14958 \pm 0.01513} & \num{0.15891 \pm 0.02089} \\
EPiC-FM (\tlep) & \num{0.10641 \pm 0.01623} & \num{0.19671 \pm 0.03365} & \num{0.18488 \pm 0.05699} & \num{1.78050 \pm 0.08955} & \num{1.48676 \pm 0.06308} \\
\midrule
\end{tabular}
\end{table*}

\begin{figure*}
    \centering
    \includegraphics[width=0.99\textwidth]{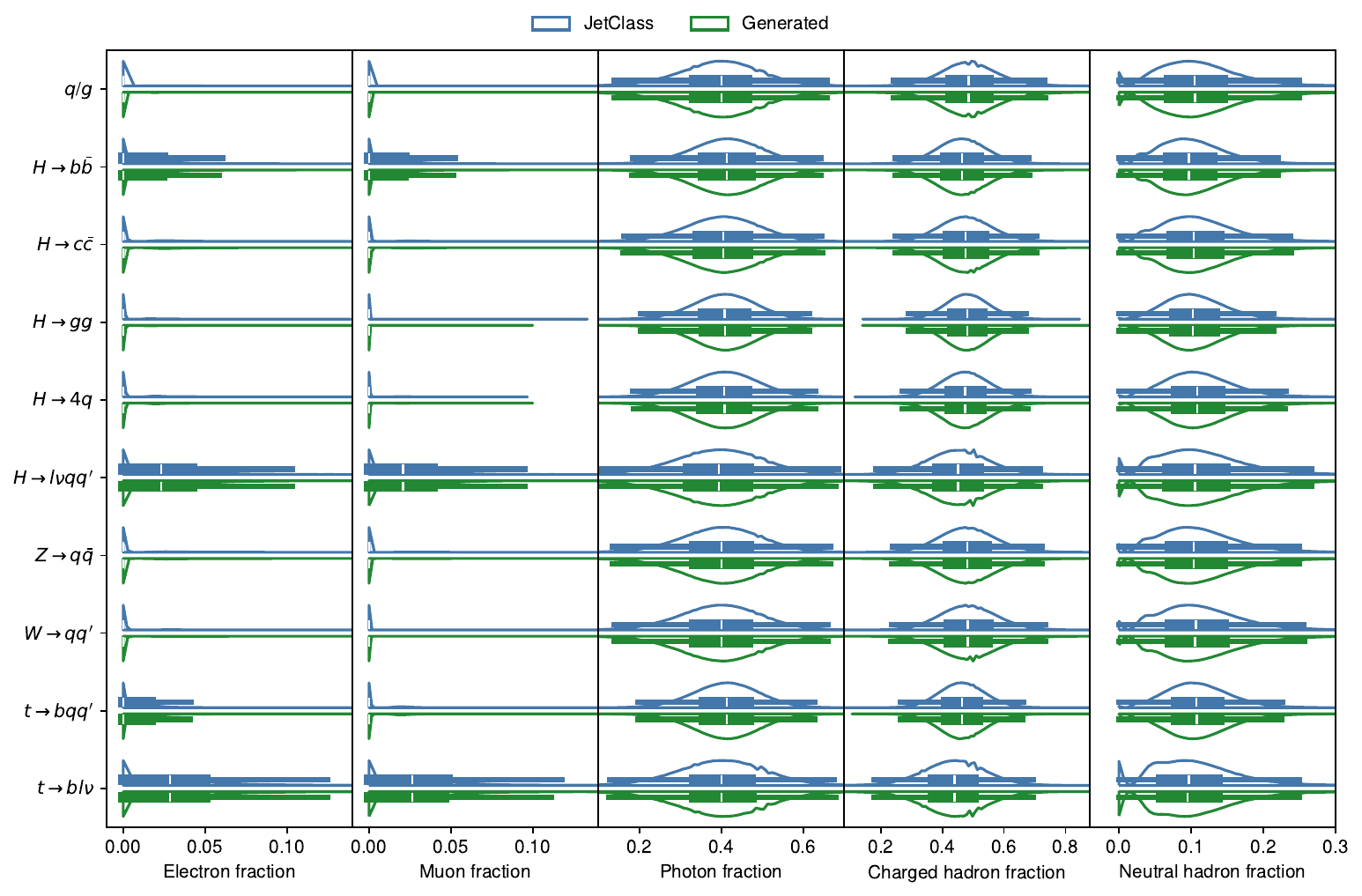}
    \caption{
        The violin plots illustrate the distribution of fractions for various
        particle types within jet constituents. The box plots show
        the median, the quartiles, and 1.5 times the interquartile range,
        with a density plot to highlight distribution peaks. Notably, the
        electron and muon fractions predominantly peak at zero across most jet
        types, resulting in box plots positioned near zero.
    }
    \label{fig:baseline_discrete_features}
\end{figure*}

A comparison of the trajectory displacement modeling is shown
in \autoref{fig:baseline_ip_and_kin_features} for \hbb, \hcc\ and \hgg\ jets.
Due to the long lifetime of $b$- and $c$-hadrons, the trajectory displacement
of jet constituents associated with those hadrons is expected to be on average
larger than for other jet constituents.
Thus, the trajectory displacement distributions of \hbb\ and \hcc\ jets
are expected to be wider than the trajectory displacement distribution
of \hgg\ jets.
Since the trajectory displacement is by definition zero for neutral particles
in the \jetclass\ dataset, only charged particles are considered in the histograms
in \autoref{fig:baseline_ip_and_kin_features}.
For all three jet types, the histograms of the generated jets agree very well
with the histograms of the real jets, showing that our model is capable of
correctly modeling the trajectory displacement of the jet constituents.
Notably, our model is able to catch the essential differences between the
trajectory displacement distributions of \hbb, \hcc\ and \hgg\ jets, which
is an important feature from the physics point of view.
However, as seen both in the ratio panels in \autoref{fig:baseline_ip_and_kin_features}
and in the corresponding KL divergence values in 
\autoref{tab:baseline_model_kld_particle_level}, the agreement between the
target distribution and the distribution obtained from the generated jets
is worse for the impact parameter features than for the kinematic features,
showing that the modeling of these distributions is more challenging. 
This could be further optimized in future work by choosing a different 
preprocessing for the  impact parameter features, by e.g. transforming them using the 
hyperbolic tangent function, which would remove the large tails of the distributions.

\begin{figure*}
    \raggedleft
    \begin{subfigure}[t]{0.99\textwidth}
        \raggedleft
        \includegraphics[width=0.903\textwidth]{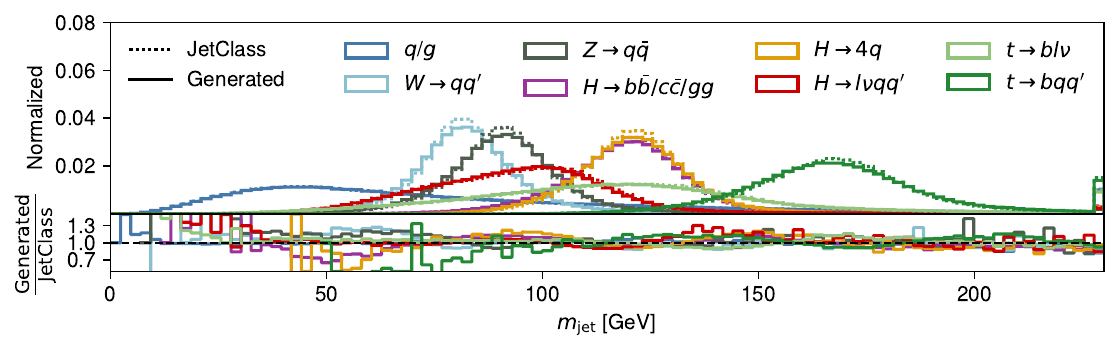} \hspace*{1cm}
        \vspace{-0.2cm}
        \subcaption{}
    \end{subfigure}
    \vspace{0.3cm}
    \begin{subfigure}[t]{0.97\textwidth}
        \raggedleft
        \includegraphics[width=0.92\textwidth]{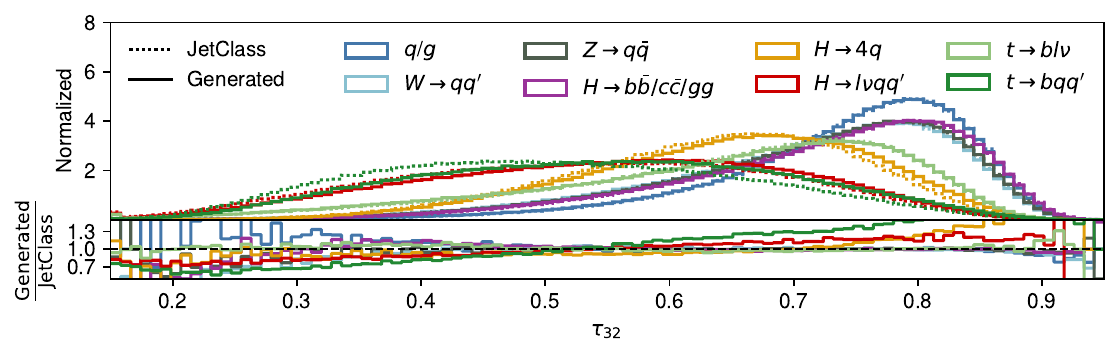} \hspace*{1cm}
        \vspace{-0.2cm}
        \subcaption{}
    \end{subfigure}
    \vspace{0.3cm}
    \begin{subfigure}[t]{0.97\textwidth}
        \raggedleft
        \includegraphics[width=0.92\textwidth]{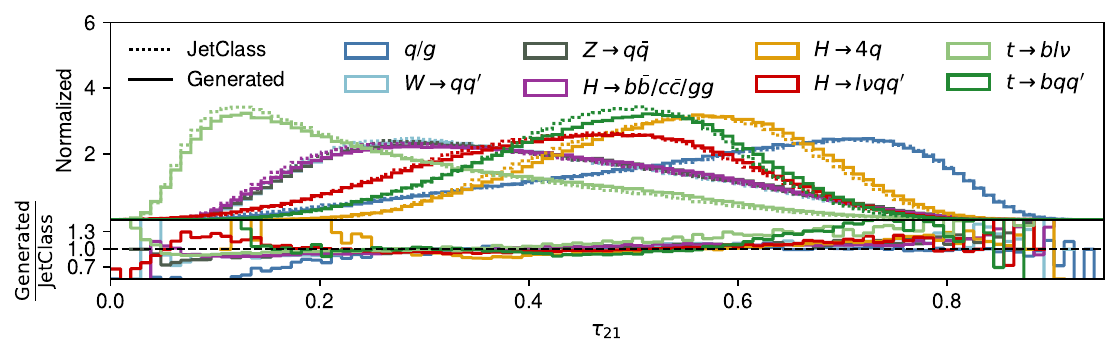} \hspace*{1cm}
        \vspace{-0.2cm}
        \subcaption{}
    \end{subfigure}
    \vspace{0.3cm}
    \begin{subfigure}[t]{0.97\textwidth}
        \raggedleft
        \includegraphics[width=0.92\textwidth]{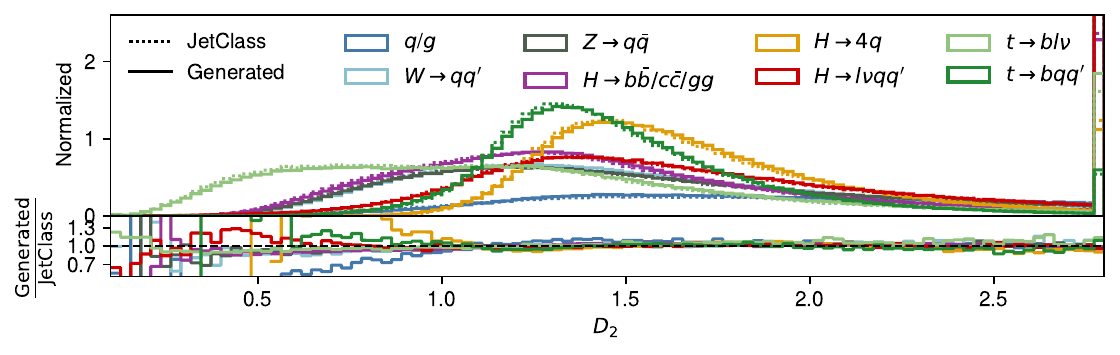} \hspace*{1cm}
        \vspace{-0.2cm}
        \subcaption{}
    \end{subfigure}
    \caption{
        Jet mass \textbf{(a)} and subjettiness ratios $\tau_{32}$ 
        \textbf{(b)}, $\tau_{21}$ \textbf{(c)} and $D_2$ \textbf{(d)} for all ten jet types.
        The histograms in dotted lines show the distributions obtained
        from the \jetclass\ dataset (i.e. real jets) and the solid lines
        show the histograms obtained from the generated jets.
        Jets from the categories \hbb, \hcc\ and \hgg\ are grouped into
        one joint histogram for better readability, since the individual
        histograms show very similar shapes.
        All histograms include underflow and overflow bins.
    }
    \label{fig:baseline_model_all_types_mass_and_tau32}
\end{figure*}
\begin{table*}
    \centering
    \caption{
        Kullback-Leibler divergence of jet-level observables and Fr\'echet physics distance with respect to the
        \jetclass\ dataset.
    }
    \label{tab:baseline_model_kld_jet_level}
\begin{tabular}{l|c|c|c|c|c}
&$\mathrm{KL}^{m_\mathrm{jet}} ~ (\times 10^{-3})$ & $\mathrm{KL}^{\tau_{21}} ~ (\times 10^{-3})$ & $\mathrm{KL}^{\tau_{32}} ~ (\times 10^{-3})$ & $\mathrm{KL}^{D_2} ~ (\times 10^{-3})$ & $\mathrm{FPD} ~ (\times 10^{-5})$\\
\midrule
Truth (\qcd) & \num{1.96024 \pm 0.22032} & \num{1.92403 \pm 0.26225} & \num{2.01218 \pm 0.27953} & \num{1.91381 \pm 0.39724} & \num{7.33742 \pm 3.30663} \\
EPiC-FM (\qcd) & \num{2.02494 \pm 0.32759} & \num{4.62940 \pm 0.47603} & \num{2.27362 \pm 0.25510} & \num{4.99645 \pm 0.47051} & \num{7.91040 \pm 1.94646} \\
\midrule
Truth (\hbb) & \num{1.92982 \pm 0.20477} & \num{1.94963 \pm 0.21244} & \num{2.08080 \pm 0.22548} & \num{1.97710 \pm 0.22072} & \num{9.98274 \pm 5.22338} \\
EPiC-FM (\hbb) & \num{3.24668 \pm 0.46254} & \num{3.63796 \pm 0.57295} & \num{2.68246 \pm 0.34426} & \num{3.35243 \pm 0.53867} & \num{12.07670 \pm 6.57340} \\
\midrule
Truth (\hcc) & \num{2.04281 \pm 0.31760} & \num{2.00572 \pm 0.22876} & \num{2.17238 \pm 0.33699} & \num{1.90518 \pm 0.17937} & \num{9.60921 \pm 4.65058} \\
EPiC-FM (\hcc) & \num{4.75491 \pm 0.45220} & \num{6.43288 \pm 0.61917} & \num{3.05689 \pm 0.48154} & \num{3.76940 \pm 0.47839} & \num{14.32434 \pm 7.91750} \\
\midrule
Truth (\hgg) & \num{1.96830 \pm 0.19475} & \num{1.83727 \pm 0.26071} & \num{1.88056 \pm 0.23067} & \num{1.97745 \pm 0.18155} & \num{8.65053 \pm 6.80698} \\
EPiC-FM (\hgg) & \num{3.74593 \pm 0.40721} & \num{3.84047 \pm 0.59242} & \num{2.80906 \pm 0.43111} & \num{3.24857 \pm 0.50123} & \num{13.46076 \pm 6.72046} \\
\midrule
Truth (\hqqqq) & \num{2.21949 \pm 0.37219} & \num{2.09503 \pm 0.30231} & \num{1.99994 \pm 0.26632} & \num{1.86540 \pm 0.31324} & \num{6.77410 \pm 4.52764} \\
EPiC-FM (\hqqqq) & \num{5.17950 \pm 0.48399} & \num{5.07132 \pm 0.47784} & \num{7.51323 \pm 0.66837} & \num{3.23005 \pm 0.47515} & \num{6.72770 \pm 1.95729} \\
\midrule
Truth (\hlnuqq) & \num{1.89667 \pm 0.30666} & \num{1.93986 \pm 0.26766} & \num{2.03454 \pm 0.23668} & \num{1.98809 \pm 0.20659} & \num{10.00879 \pm 6.87615} \\
EPiC-FM (\hlnuqq) & \num{3.48160 \pm 0.48279} & \num{3.45981 \pm 0.43930} & \num{9.25649 \pm 0.88029} & \num{2.31713 \pm 0.33259} & \num{13.79823 \pm 4.53244} \\
\midrule
Truth (\zqq) & \num{2.06754 \pm 0.17644} & \num{1.99159 \pm 0.32238} & \num{1.95261 \pm 0.23823} & \num{2.04952 \pm 0.19436} & \num{7.52867 \pm 3.61417} \\
EPiC-FM (\zqq) & \num{4.81740 \pm 0.54586} & \num{3.52108 \pm 0.46385} & \num{1.96704 \pm 0.21247} & \num{2.44368 \pm 0.25113} & \num{10.14084 \pm 5.14489} \\
\midrule
Truth (\wqq) & \num{2.05795 \pm 0.39753} & \num{2.04785 \pm 0.22169} & \num{2.03512 \pm 0.19298} & \num{1.99837 \pm 0.24655} & \num{7.66854 \pm 5.33720} \\
EPiC-FM (\wqq) & \num{5.74197 \pm 0.43918} & \num{4.26782 \pm 0.63239} & \num{2.67976 \pm 0.41067} & \num{3.15276 \pm 0.49860} & \num{12.76336 \pm 7.56694} \\
\midrule
Truth (\thad) & \num{1.86110 \pm 0.25434} & \num{2.00490 \pm 0.26157} & \num{2.06202 \pm 0.24148} & \num{1.94072 \pm 0.31298} & \num{7.45729 \pm 3.61879} \\
EPiC-FM (\thad) & \num{4.67384 \pm 0.60934} & \num{7.52936 \pm 0.76181} & \num{30.02929 \pm 1.07748} & \num{2.95281 \pm 0.43531} & \num{9.22545 \pm 3.82906} \\
\midrule
Truth (\tlep) & \num{2.00893 \pm 0.14388} & \num{1.85344 \pm 0.22582} & \num{1.98487 \pm 0.21507} & \num{2.00167 \pm 0.27673} & \num{8.11577 \pm 5.27167} \\
EPiC-FM (\tlep) & \num{3.38949 \pm 0.23017} & \num{5.66341 \pm 0.49231} & \num{2.56432 \pm 0.36011} & \num{3.75036 \pm 0.46000} & \num{7.96196 \pm 4.10444} \\
\midrule
\end{tabular}
\end{table*}

The evaluation of the particle-ID modeling is shown in
\autoref{fig:baseline_discrete_features}, where we show violin plots and
box plots for the fraction of jet constituents of different particle types for
all ten jet types.
The agreement between the generated jets and the real jets is very good for
all jet types, showing that on average the generated jets contain the same
fraction of different particle types as the real jets.
The modeling of the electric charge also shows very good agreement, which
is shown in Appendix \ref{sec:charge_modeling}.

\subsection{Jet substructure modeling}

The jet mass $m_\mathrm{jet}$, the two subjettiness ratios $\tau_{32}$ and
$\tau_{21}$, and $D_2$ are shown in \autoref{fig:baseline_model_all_types_mass_and_tau32}
for the different jet types.
The real jets are shown in dotted lines while the generated jets
are shown in solid lines.
Both the jet mass and the subjettiness ratio distributions show very good agreement
for all jet types.
The largest deviations between the target distribution and the distribution
of the generated jets are seen for \thad\ and \hlnuqq , where the distribution of the generated
jets peaks at a larger value of $\tau_{32}$.
This mismodeling also shows in the values of the KL divergence which are listed in 
\autoref{tab:baseline_model_kld_jet_level} for some of the jet-level
observables.
In addition to the 1-dimensional KL divergence values, we also evaluate
the multi-dimensional distribution by calculating the Fr\'echet physics distance (FPD)
as introduced in~\cite{Kansal:2022spb}, which is also listed in 
\autoref{tab:baseline_model_kld_jet_level}.
For the FPD calculation we follow the recommendations from~\cite{Kansal:2022spb},
thus we evaluate the FPD based on a set of 36 EFPs (all EFPs of degree less than
five) and use the \texttt{JetNet}~\cite{Kansal_JetNet_2023} library
to calculate both the EFPs and the corresponding FPD.
Further studies were done to determine whether narrowing down our 
model's features to just kinematics and whether training solely 
on \thad\ jets enhances the modeling of the \thad\ substructure, 
which can be found in Appendix \ref{sec:further_studies_for_tbqq_modeling}.

\subsection{Classifier test}

\begin{figure*}
    \centering
    \begin{subfigure}[t]{0.48\textwidth}
        \includegraphics[width=0.99\textwidth]{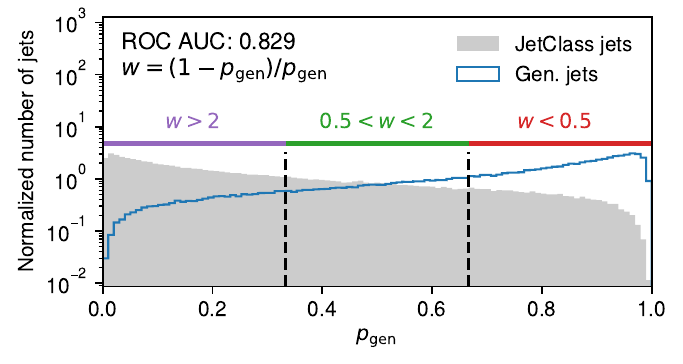}
        \subcaption{}
        \label{fig:classifier_test_output}
    \end{subfigure}
    \begin{subfigure}[t]{0.48\textwidth}
        \includegraphics[width=0.96\textwidth]{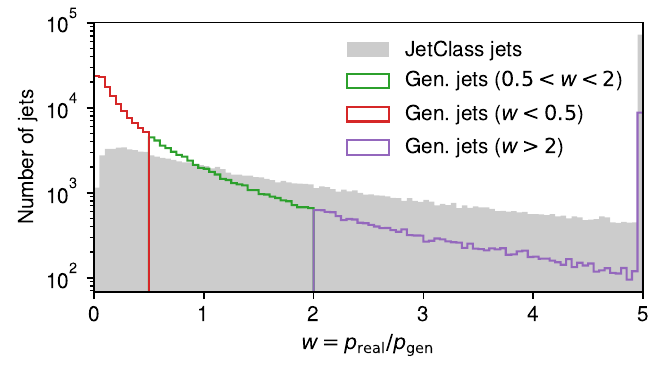}
        \subcaption{}
        \label{fig:classifier_test_weights}
    \end{subfigure}
    \begin{subfigure}[t]{0.48\textwidth}
        \includegraphics[width=0.99\textwidth]{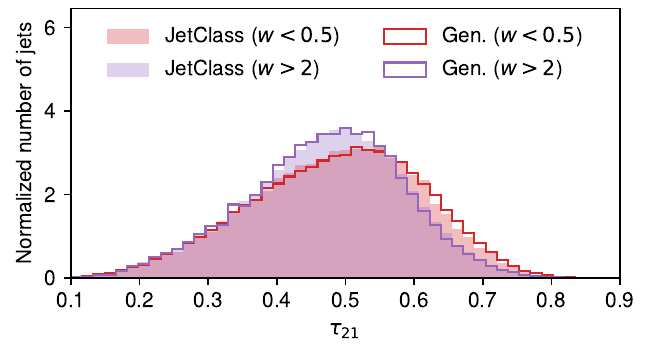}
        \subcaption{}
        \label{fig:classifier_test_tau21}
    \end{subfigure}
    \begin{subfigure}[t]{0.48\textwidth}
        \includegraphics[width=0.99\textwidth]{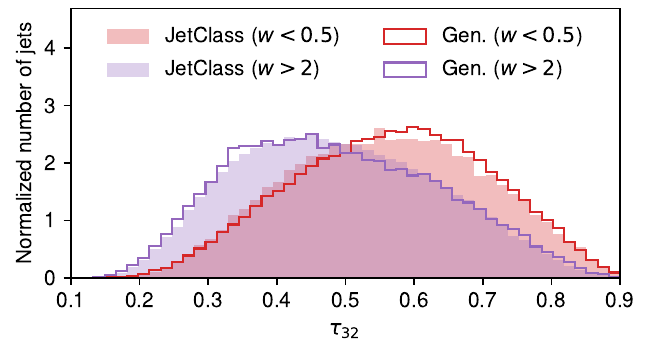}
        \subcaption{}
        \label{fig:classifier_test_tau32}
    \end{subfigure}
    \caption{
        Results of the classifier test for \thad\ jets. Figure \textbf{(a)} shows
        the classifier output for real and generated jets, as well as the
        cut values for the three regions. Figure \textbf{(b)} shows
        the weight distribution of the generated jets with differently colored
        histograms for the generated jets from the different regions. 
        Figures \textbf{(c)} and \textbf{(d)} show the $\tau_{21}$ and $\tau_{32}$
        distributions for the real jets and the generated jets from the regions where
        the classifier output distribution is dominated by real jets or generated jets.
    }
    \label{fig:classifier_test}
\end{figure*}

In addition to the evaluation presented in the previous subsections, we also
investigate the performance of our model with the classifier test proposed in Ref.~\cite{CaloFlowKrause_2023}.
Thus, a binary classifier is trained to distinguish between real jets from
the \jetclass\ dataset and generated jets that were generated with our model.
This classifier test is performed for each jet type separately.

To train the classifier, we use the 500k generated jets and 500k 
real jets from the previous evaluation, and split them into 250k, 
50k and 200k jets for training,  validation and testing of the 
classifier, respectively.
The classifier is trained by fine-tuning the ParT-kin version from Ref.~\cite{ParT},
which means that the last MLP of the ParT architecture is re-initialized and 
changed to two output nodes.
The classifier is then trained with real vs generated jets using the 
AdamW optimizer~\cite{AdamW} with a learning rate of 0.001
for a few epochs. 
However, it should be noted that the classifier training was not 
extensively optimized.
We noticed that the validation loss converges after
just a few epochs after which the classifier starts to overfit.
We therefore chose to use the model with the highest validation accuracy
before clear signs of overfitting are visible.

\begin{table}[h]
    \centering
    \caption{
        AUC scores of the classifier test for the ten different jet types.
    }
    \label{tab:classifier_test_auc_scores}
    \begin{tabular}{l c}
        \normalsize
        Jet type & AUC score\\
        \midrule
        \midrule
        \qcd & \num{0.681} \\
        \wqq & \num{0.683} \\
        \zqq & \num{0.689} \\
        \hlnuqq & \num{0.706} \\
        \tlep & \num{0.706} \\
        \hbb & \num{0.712} \\
        \hgg & \num{0.716} \\
        \hcc & \num{0.728} \\
        \hqqqq & \num{0.809} \\
        \thad & \num{0.829} \\
        \midrule
    \end{tabular}
\end{table}

The AUC values of the corresponding ROC curves are shown in 
\autoref{tab:classifier_test_auc_scores}.
To estimate the uncertainty of the classifier scores, five
classifiers are trained for \hbb\ jets with different random seeds.
All classifiers converge at similar performance, with the standard
deviation of the AUC scores being 
\mbox{$\sigma_\mathrm{AUC}(H\to b\bar{b}) = \num{0.003}$}.
The largest AUC value is obtained for \thad\ jets with an AUC
of \num{0.829}, confirming again that this is the most challenging 
jet type for our model.
The results of the corresponding classifier test are shown in 
\autoref{fig:classifier_test} in more detail.
However, the AUC value for \thad\ jets is still well below 1, 
allowing us to investigate the \thad\ jets corresponding
to different regions of the classifier output.
Therefore, following Ref.~\cite{understandLimitationsOfGen}, we split the \thad\ jets into three regions, based on their
weight $w = p_\mathrm{real} / p_\mathrm{gen}$ as illustrated in
\autoref{fig:classifier_test_output} and \autoref{fig:classifier_test_weights}.
The first region contains jets with $w > 2$, the second region contains
jets with $0.5 < w < 2$, and the third region contains jets with $w < 0.5$.
Thus, the outer regions correspond to jets that are clearly identified as
either real or generated jets, while the center region contains jets where the
classifier is not able to clearly distinguish them.

The corresponding subjettiness ratios $\tau_{21}$ and $\tau_{32}$ are shown in
\autoref{fig:classifier_test_tau21} and \autoref{fig:classifier_test_tau32}
for \thad\ jets from the outer regions.
This shows the jet substructure differences between the \thad\ jets that are clearly 
classified as either real or generated.
\thad\ jets that correspond to $w > 2$, i.e. jets of which the classifier is rather
certain that they are real jets,
have on average a smaller $\tau_{32}$ value than \thad\ jets that are clearly identified
as generated jets ($w < 0.5$).
Therefore, the classifier classifies \thad\ jets that are more 3-prong-like
(i.e. with a smaller $\tau_{32}$ value) as real \thad\ jets, 
while \thad\ jets that are less 3-prong-like (i.e. with a larger $\tau_{32}$
value) tend to be classified as generated \thad\ jets.
Thus, our model seems to struggle with the modeling of the 3-prong structure of
\thad\ jets, which is also consistent with the results from the previous
sections, where we saw that the $\tau_{32}$ distribution of the generated
\thad\ jets peaks at a slightly larger value than the $\tau_{32}$ distribution
of the real \thad\ jets.
The peak at the larger $\tau_{32}$ value and thus the overall shift of
the distribution is caused by the poorly modeled jets:
given that \SI{61}{\percent} of the generated \thad\ jets have $w < 0.5$, 
the overall $\tau_{32}$ distribution of the generated jets is dominated by
jets with larger $\tau_{32}$.
On the other hand, only \SI{15}{\percent} of the real \thad\ jets have $w < 0.5$,
and \SI{58}{\percent} of the real \thad\ jets have $w > 2$, which correspond
to the purple distribution in \autoref{fig:classifier_test_tau32}.

\section{Conclusion}

\label{sec:conclusion}

Generative surrogate models of different physical processes are a powerful new tool with multiple potential applications.
So far, speeding up classical simulations, either of particle interactions in calorimeters or of the calculation of matrix elements, has been the primary focus of research.
However, learning surrogate models of jets has its own interesting aspects:
i) such models can, in principle, be trained on collider data, allowing the interpolation of distributions to estimate backgrounds;
ii) they can form parts of a fully differentiable analysis and interpretation chain;
iii) an interpretable version trained directly on collider data may provide an additional way of gaining insight into showering and hadronization processes;
iv) learning jets at the constituent level is a crucial stepping stone to simulating full events at the finest granularity;
and v) they provide a meaningful testbed to benchmark progress in generative models for point clouds.

The main result of this work is that the EPiC-FM~\cite{epicly} framework can be easily scaled up to more complex datasets. The move from the \textsc{JetNet} to the \jetclass\ dataset results in more classes (10 instead of 5), more features per jet (13 instead of 3), geometrically larger jets (\mbox{$R=0.8$} instead of \mbox{$R=0.4$}), and finally more jets available for training with \jetclass\ containing a factor of 70 more jets per jet type than \textsc{JetNet}.

Overall we observe an accurate description of physical quantities. For some individual features and classes, the agreement between generated data and ground truth is on the same level as the statistical fluctuations between different samples drawn from the ground truth distribution. In general, more complex final states (i.e. multi-pronged signals with intermediate resonances) are more difficult to describe than relatively simple light quark or gluon-induced jets.

The capability to model quantities beyond jet kinematics demonstrated here will serve as an important factor in further increasing the performance of anomaly detection algorithms. Previous results already showed that going from a few high-level to many low-level features increases the maximum achievable significance improvement by more than a factor of three~\cite{full_phase_space_anomaly}. 
Access to trajectory displacement and particle identification quantities now enable these models to be sensitive to final states including $b$-flavor and for example allow automatically covering long-lived~\cite{long_lived_big_review} final states with anomaly detection as well, greatly increasing their accessible theory phase space.

Finally, the \jetclass\ dataset has sufficient complexity to serve as a useful next benchmark for generative models applied to point-cloud-like data. 
Accordingly, we provide an extensive set of one-dimensional (KL-divergences) and multi-dimensional (classifier scores) evaluation methods for future reference and comparison.

\section*{Code Availability}
The code used to produce the results presented in this paper is available at 
\url{https://github.com/uhh-pd-ml/beyond_kinematics}.

\section*{Acknowledgements}
The authors thank Huilin Qu for his support with the \jetclass\ dataset
and helping with any questions.
EB is funded by a scholarship of the Friedrich Naumann Foundation for Freedom and by the German Federal Ministry of Science and Research (BMBF) via Verbundprojekts 05H2018 - R\&D COMPUTING (Pilotmaßnahme ErUM-Data) Innovative Digitale Technologien für die Erforschung von Universum und Materie.
JB, EB, CE, and GK acknowledge support by the Deutsche Forschungsgemeinschaft under Germany’s Excellence Strategy – EXC 2121  Quantum Universe – 390833306 
and via the KISS consortium (05D23GU4) funded by the German Federal Ministry of Education and Research BMBF in the ErUM-Data action plan.
DS is supported by DOE grant DOE-SC0010008. 
This research was supported in part through the Maxwell computational resources operated at Deutsches Elektronen-Synchrotron DESY, Hamburg, Germany.

\bibliography{refs}

\appendix

\newpage

\section{Hyperparameters}
\label{sec:appendix_hyperparameters}

The hyperparameters used for our generative model are listed in
\autoref{tab:baseline_model_hyperparameters}.
This choice of hyperparameters was found to lead to the best performance
among several different configurations, though they were not
extensively optimized.

\begin{table}[h]
    \caption{
        Hyperparameters used for the baseline EPiC-FM model.
    }
    \label{tab:baseline_model_hyperparameters}
    \begin{tabular}{l r}
        Hyperparameter & Value \\
        \midrule
        \midrule
        Number of EPiC layers & 20 \\
        Global dim. in EPiC layers & 16 \\
        Hidden dim. in EPiC layers & 300 \\
        ODE solver & midpoint \\
        Number of function evaluations & 200 \\
        Activation function & LeakyReLU(0.01) \\
        AdamW~\cite{AdamW} learning rate & 0.001 \\
        Weight decay & 0.00005 \\
        Learning rate scheduler & CosineAnnealing \\
        Linear warm-up & 20 epochs \\
        Total number of epochs & EPiC-FM: 500 \\
        & EPiC-FM-kin: 500 \\
        & EPiC-FM-top: 1000 \\
        & EPiC-FM-kintop: 1000 \\
        Batch size & 1024 \\
        Trainable parameters & 8,504,698 \\
        Training jets & 3M (300k per jet type)\\
        \midrule
    \end{tabular}
\end{table}

The hyperparameters used for the classifier test are shown in
\autoref{tab:classifier_hyperparameters}.
The model weights are initialized using the pre-trained
ParT-kin model from \url{https://github.com/jet-universe/particle_transformer}.

\begin{table}[h]
    \caption{
        Hyperparameters used for the classifier test with the
        ParT-kin~\cite{ParT} model.
    }
    \label{tab:classifier_hyperparameters}
    \begin{tabular}{l r}
        Hyperparameter & Value \\
        \midrule
        \midrule
        AdamW~\cite{AdamW} learning rate & 0.001 \\
        Weight decay & 0.00005 \\
        Learning rate scheduler & Constant \\
        Total number of epochs & 100 \\
        Batch size & 256 \\
        Trainable parameters & \num{2141134} \\
        Training jets & \num{500000} \\
        Validation jets & \num{100000} \\
        Testing jets & \num{400000} \\
        \midrule
    \end{tabular}
\end{table}

\section{Modeling of the electric charge}
\label{sec:charge_modeling}

\begin{figure}[h]
    \centering
    \includegraphics[width=0.99\linewidth]{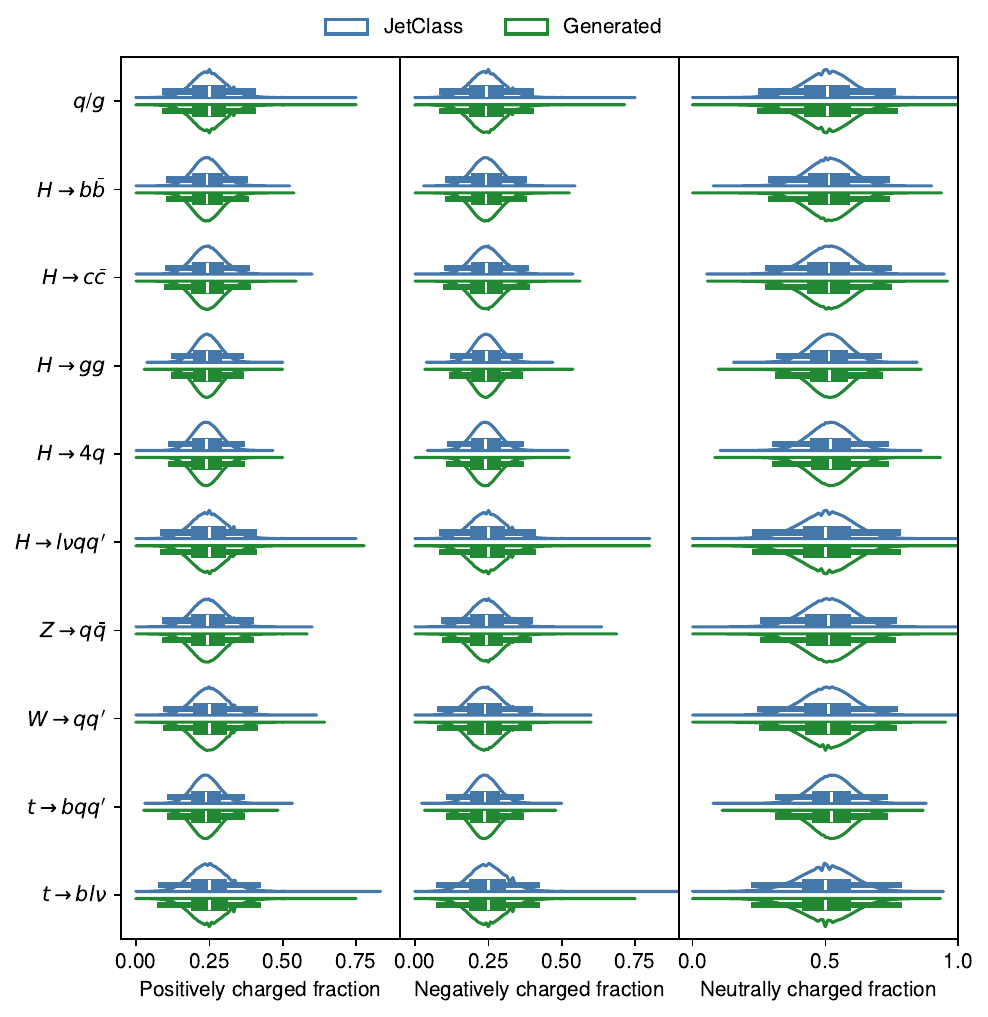}
    \caption{
        The violin plots depict the distribution of the fraction of jet
        constituents with different electric charges. Each panel includes a box
        plot that highlights the median, quartiles, and 1.5 times the
        interquartile range.
    }
    \label{fig:baseline_charge}
\end{figure}

Similar to the evaluation of the particle-ID features,
we show the distribution and the box plots for the fraction of particles
corresponding
to different electric charge values in \autoref{fig:baseline_charge}.

While the spread is quite large as indicated by the
intervals in \autoref{fig:baseline_charge}, around half of
the jet constituents are electrically neutral and correspondingly
around a quarter of the constituents carries charge $-1$ and $1$,
respectively.

\begin{figure*}
    \centering
    \begin{subfigure}[t]{0.48\textwidth}
        \raggedright
        \includegraphics[height=7cm]{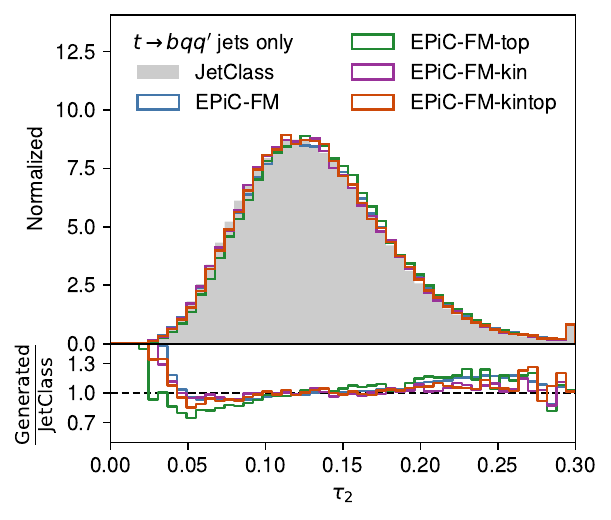}
    \end{subfigure}\hfill
    \begin{subfigure}[t]{0.49\textwidth}
        \raggedright
        \includegraphics[height=7cm]{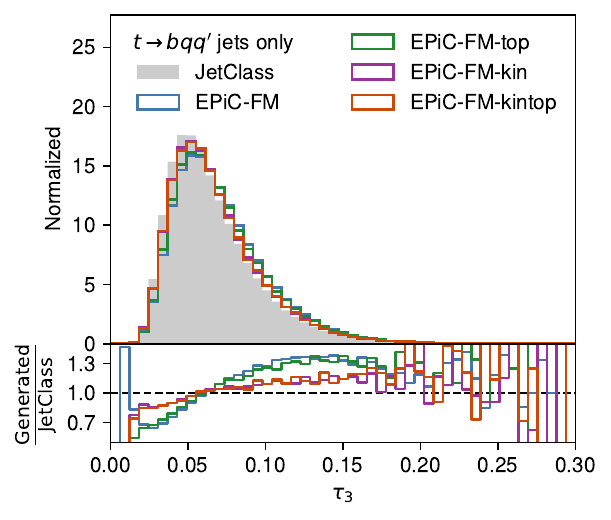}
    \end{subfigure}
    \begin{subfigure}[t]{0.48\textwidth}
        \raggedright
        \includegraphics[height=7cm]{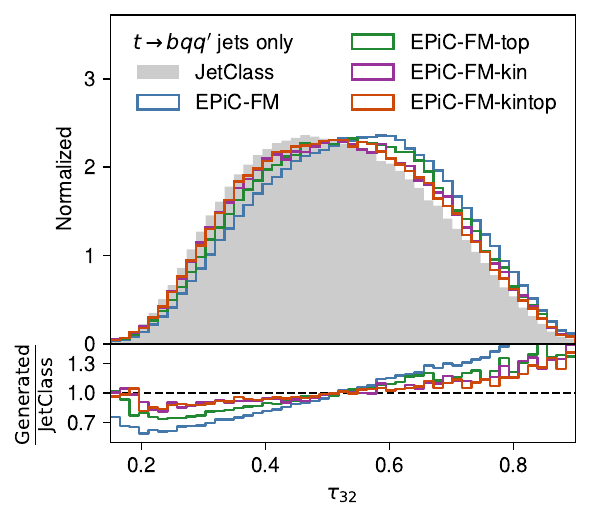}
    \end{subfigure}\hfill
    \begin{subfigure}[t]{0.49\textwidth}
        \raggedright
        \includegraphics[height=7cm]{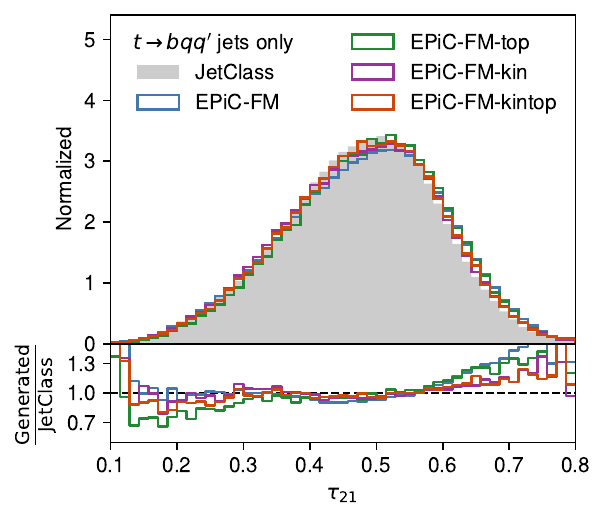}
    \end{subfigure}
    \begin{subfigure}[b]{0.49\textwidth}
        \raggedright
        \includegraphics[height=7cm]{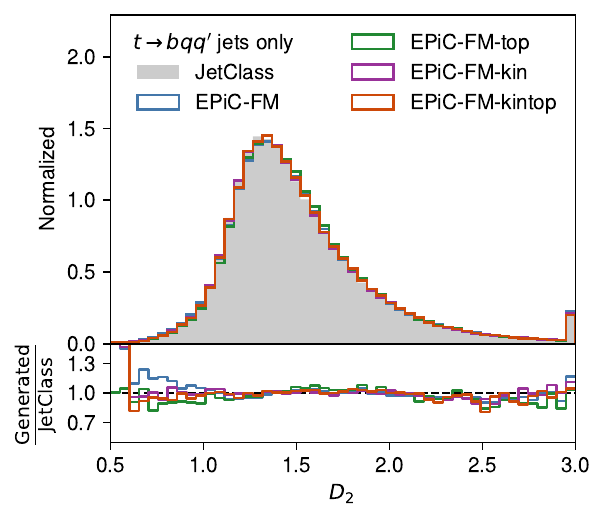}
    \end{subfigure}\hfill
    \begin{subfigure}[b]{0.49\textwidth}
        \raggedright
        \includegraphics[height=7cm]{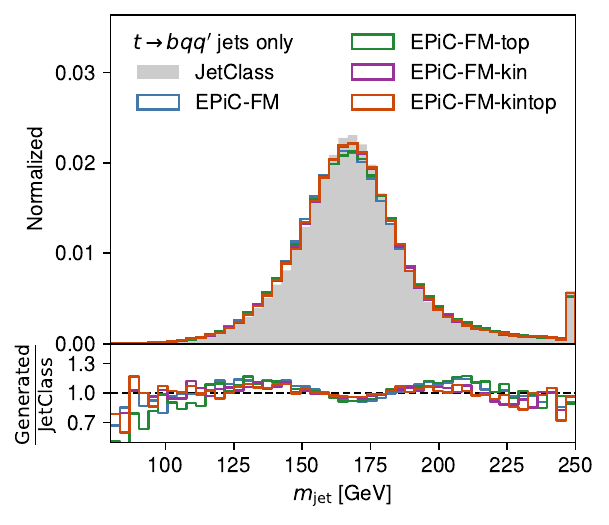}
    \end{subfigure}
    \caption{
        Comparison of the jet substructure of \thad\ jets for the four different models.
        The models EPiC-FM-top and EPiC-FM-kintop were trained on \thad\ jets only.
    }
    \label{fig:ablation_studies_substructure_comparison}
\end{figure*}

\begin{table*}[t]
    \caption{
      KL divergence metrics and FPD values for the four different models used for improving the modeling
        of \thad\ jets.
      The models EPiC-FM-top and EPiC-FM-kintop were trained on \thad\ jets only.
    }
    \label{tab:baseline_vs_kin_vs_kintop_vs_alltop}
\begin{tabular}{l|c|c|c|c|c}
&$\mathrm{KL}^{m_\mathrm{jet}} ~ (\times 10^{-3})$ & $\mathrm{KL}^{\tau_{21}} ~ (\times 10^{-3})$ & $\mathrm{KL}^{\tau_{32}} ~ (\times 10^{-3})$ & $\mathrm{KL}^{D_2} ~ (\times 10^{-3})$ & $\mathrm{FPD} ~ (\times 10^{-5})$\\
\midrule
Truth (\thad) & \num{1.86110 \pm 0.25434} & \num{2.00490 \pm 0.26157} & \num{2.06202 \pm 0.24148} & \num{1.94072 \pm 0.31298} & \num{7.45729 \pm 3.61879} \\
EPiC-FM (\thad) & \num{4.67384 \pm 0.60934} & \num{7.52936 \pm 0.76181} & \num{30.02929 \pm 1.07748} & \num{2.95281 \pm 0.43531} & \num{9.22545 \pm 3.82906} \\
EPiC-FM-kin (\thad) & \num{2.67246 \pm 0.25860} & \num{3.13963 \pm 0.37574} & \num{6.15219 \pm 0.77443} & \num{2.10880 \pm 0.33337} & \num{8.04998 \pm 2.76330} \\
EPiC-FM-top (\thad) & \num{4.34749 \pm 0.44568} & \num{7.60276 \pm 0.58837} & \num{13.35373 \pm 0.89500} & \num{3.16551 \pm 0.25745} & \num{11.14224 \pm 3.72043} \\
EPiC-FM-kintop (\thad) & \num{2.57447 \pm 0.38647} & \num{3.48134 \pm 0.31294} & \num{4.81889 \pm 0.63065} & \num{2.33203 \pm 0.34413} & \num{8.10681 \pm 3.00268} \\
\midrule
\end{tabular}
\end{table*}

\section{Further studies for hadronic top jet modeling}
\label{sec:further_studies_for_tbqq_modeling}

Since \thad\ jets were found to be the most challenging jet type to model, further
studies are performed to improve and investigate the modeling of those jets.
To this end, we compare our baseline model, which 
we refer to in the following as EPiC-FM, to three additional models:
\begin{itemize}
\item \textbf{EPiC-FM-top} is trained on the same features as
        the baseline model, but only the \thad\ jets from the training dataset
        are used for training;
\item \textbf{EPiC-FM-kin} is trained on kinematic features only 
        (i.e. \ptrel, \etarel\ and \phirel) but uses the full training
        dataset and 
\item \textbf{EPiC-FM-kintop} is trained on kinematic features only 
        (i.e. \ptrel, \etarel\ and \phirel) and only uses the \thad\ jets 
        from the training dataset.
\end{itemize}

The KL divergence values for the jet-level observables are listed for all
models in \autoref{tab:baseline_vs_kin_vs_kintop_vs_alltop} and
the substructure of the generated \thad\ jets for the three different models is shown
in \autoref{fig:ablation_studies_substructure_comparison}.
It can be seen that the models that are restricted to kinematic features
show better agreement with the real jets than the two models that are trained
on all features.
This is also confirmed by the KL divergence, which show that the EPiC-FM-kin
model has a significantly smaller KL divergence than the EPiC-FM model 
for all substructure observables.
These observations lead to the conclusion that the additional features slightly 
confuse the modeling of the kinematic features which results in a worse 
representation of the substructure. However, the jets are still modeled 
well in the case of all features, and the ability to model the discrete 
features extends the usability of such generative models.

The model that is restricted to \thad\ jets and kinematic features, 
EPiC-FM-kintop, shows similar improvements as the EPiC-FM-kin model. 
However, while there is a slight improvement over the EPiC-FM-kin model
in the KL divergence for $\tau_{32}$, the KL divergence
for $\tau_{21}$ and $D_2$ is slightly larger than for the kin-only model.
In the case where all features are used, i.e. when comparing the EPiC-FM-top model
to the baseline model EPiC-FM, we see an improvement in $\tau_{32}$, while 
not seeing an improvement in the KL divergence of $\tau_{21}$ and $D_2$.
Thus, the single jet-type model performs better than the model that is trained
on all jet types if all features are used, while there is no clear improvement 
by training solely on \thad\ jets if only kinematic features are used.
All models show similar performance in the Fr\'echet physics distance (FPD) metric.

\end{document}